\documentclass[a4paper,12pt]{article}

\usepackage{mathptmx}
\usepackage{setspace}

\usepackage[left=2.5cm,right=2.5cm,top=2.5cm,bottom=2.5cm]{geometry}

\usepackage{titlesec}
\titleformat{\section}{\bfseries\large}{\thesection\ }{1em}{}
\titleformat{\subsection}{\bfseries}{\thesubsection\ }{1em}{}

\usepackage[hidelinks]{hyperref}

\usepackage{datetime}
\newdateformat{mydate}{\THEYEAR-\twodigit{\THEMONTH}-\twodigit{\THEDAY}}

\usepackage[sort&compress,numbers]{natbib}
\usepackage{abstract}
\usepackage{titling}
\usepackage{graphicx}
\usepackage{subfig}

\begin{document}

\begin{center}
    \vspace{2cm}
    {\Large\bfseries 
A method for converting high energy physics detector description into a Unity visualization \par}
    \vspace{0.5cm}
    {\large (Presented at the 32nd International Symposium on Lepton Photon Interactions at High Energies\par} 
    {\large Madison, Wisconsin, USA, August 25-29, 2025) \par}
    \vspace{0.3cm}
    {\large Tianzi Song$^{1}$, Yumei Zhang$^{2}$, Zhengyun You$^{1}$\par}
    \vspace{0.2cm}
    {\small
    $^1$ School of Physics, Sun Yat-sen University, Guangzhou, 510275, China \\
    $^2$ Sino-French Institute of Nuclear Engineering and Technology, Sun Yat-sen University, Zhuhai, 519082, China \\
    }
    {\small Email: songtz@mail2.sysu.edu.cn, zhangym26@mail.sysu.edu.cn, youzhy5@mail.sysu.edu.cn}
    \vspace{0.3cm}
    {\mydate\today}
\end{center}

\vspace{0.4cm}

\begin{abstract}
\noindent
Detector visualization plays a vital role 
in high energy physics (HEP) experiments, yet existing detector descriptions, such as GDML, lack compatibility with industrial 3D tools. 
We present an automated conversion framework that transforms four major HEP detector descriptions, including GDML, Geant4, ROOT and DD4hep, into standardized FBX models compatible with a industrial 3D platform called Unity. 
This solution enables HEP detectors to be directly visualized in the professional 3D ecosystem, which is of great help for detector design verification, event display development, and public participation. 

\end{abstract}

\vspace{0.6cm}

\section{Introduction}
\label{sec:introduction}
Detector visualization plays a critical role in modern HEP experiments, enabling intuitive support for design, assembly, commissioning, and physics analysis by serving as the foundation for event display~\cite{BESIII analysis}. 
Currently, specialized geometry frameworks, such as Geant4~\cite{Geant4}, ROOT~\cite{ROOT}, Geometry Description Markup Language~(GDML)~\cite{GDML}, and the Detector Description Toolkit for HEP~(DD4hep)~\cite{DD4hep}, are employed to model intricate detector geometries. 
However, these tools are largely confined to Linux environments and offer visualization features that are 
difficult to use. 
Furthermore, the HEP field lacks a unified visualization solution that integrates advanced techniques and supports multiple platforms. 

Unity, a professional engine for interactive content creation, stands out for its powerful visualization capabilities and broad system compatibility. 
It has been successfully implemented in experiments like JUNO, BESIII, and Belle II for detector visualization~\cite{BESIII unity, JUNO unity}. 
Yet, direct import of geometry descriptions from existing frameworks into Unity remains unfeasible. 
To bridge this gap, we developed an interface~\cite{mygithub, unity} based on the HEP Software Foundation (HSF) geometry writer~\cite{HSF Writer}.
This interface can operate within Geant4 to automatically convert detector descriptions into the FilmBox~(FBX) format, a widely adopted 3D interchange standard format that can be seamlessly imported into Unity. 
We evaluated this interface by applying it to four HEP detectors, each utilizing a distinct geometric description format.

\section{Detector descriptions}
\label{sec:description}
Presently, four established tools are mainly used for detector descriptions in HEP field, notably Geant4, ROOT, GDML, and DD4hep. 
Geant4~\cite{Geant4} is a leading toolkit for modeling particle-matter interactions, incorporating a dedicated geometry system to define detector structures. 
ROOT~\cite{ROOT}, in contrast, functions as a multi-purpose data analysis framework widely employed in HEP. 
GDML~\cite{GDML}, an XML-based language, offers a standardized and platform-agnostic method for encoding detector configurations. 
Meanwhile, DD4hep~\cite{DD4hep} presents an integrated solution that
unifies simulation, reconstruction, and analysis workflows. 

Unity serves as a professional engine for video and game development, enabling effective visualization functions. 
Numerous HEP experiments have integrated Unity into their offline software systems.
Visualization applications built on Unity provide several key benefits:
\begin{itemize}
    \item \emph{Advanced 3D display capabilities.} 
    Unity benefits from sustained financial and technical investment, resulting in leading-edge 3D rendering and graphics capabilities.
    \item \emph{Extensive platform support.} 
    Unity is compatible with more than 20 platforms, offering broad hardware adaptability and seamless operation across different environments. 
    \item \emph{No-cost licensing.} 
    For non-commercial endeavors such as HEP experiments, Unity is provided free of charge.
    \item \emph{Enhanced extensibility.} 
    Unity supports advanced software development, allowing for future expansions like event display features. 
\end{itemize}

\section{Methodologies}
\label{sec:method}
Despite its numerous advantages for detector visualization, Unity also presents several challenges. 
HEP experiments utilize highly complex detectors with billions of components, each supported by dedicated detector description systems in their offline software. 
The frequent updates and maintenance required for these descriptions make manual reconstruction of detector geometry in Unity both time-intensive and error-prone.
To address this, we have presented an interface~\cite{mygithub} that automatically converts detector descriptions from Geant4, ROOT, GDML, and DD4hep directly into Unity. 

To ensure geometric consistency across applications, HEP offline software typically employs multiple interfaces that adapt a unified detector description source for various uses. 
For instance, an existing GDML detector description can be automatically processed for construction in Geant4 and ROOT via the GDML–Geant4 and GDML–ROOT interfaces~\cite{ROOT_Geant4}, respectively. 
This work leverages similar interfaces, such as ROOT-GDML and DD4hep-GDML, to streamline the consolidation of detector description into GDML format or their direct implementation in Geant4. 
Utilizing the HSF Geometry Writer, we have established a method for converting detector data from GDML to Unity~\cite{mygithub}. 
This process systematically translates the HEP detector description and its constituent elements—such as material lists, positional and rotational data, and physical nodes—into the FBX file format, maintaining consistency throughout.

Earlier techniques for converting detector description, including the GDML-FreeCAD-Pixyz-FBX chain~\cite{BESIII unity} or direct DD4hep to FBX conversion~\cite{DD4hep_unity}, often faced issues like extended conversion pipelines or constrained format adaptability. 
The interface proposed in this study delivers several improvements:
\begin{itemize}
    \item \emph{Comprehensive Coverage.} 
    It supports all four detector descriptions, 
    including Geant4, ROOT, GDML, and DD4hep.
    \item \emph{Efficient Workflow.} 
    The conversion to FBX is achieved in a single step, with processing times typically within minutes, accommodating even highly complex detectors.
    \item \emph{Adaptable Configuration.} 
    Geometric shape precision can be directly adjusted within Geant4, offering greater versatility.
    \item \emph{Accessible Resources.} 
    All software involved in the conversion are free or open source.
    \item \emph{Superior Geometric Support.} 
    Geant4 encompasses a wider variety of geometry primitives than DD4hep, enhancing its capability to manage intricate detector conversions.
\end{itemize}

\section{Visualization in Unity}
\label{sec:visualization}
To assess the interface's compatibility with four distinct detector description formats, we processed detectors based on each format into FBX and imported them into Unity. 

\subsection{From Geant4 to Unity with JUNO detector}
\label{subsec:JUNO}
\begin{figure}[ht]
  \centering 
  \subfloat[JUNO detector]{
        \label{fig:JUNO}
        \includegraphics[width=0.36\textwidth]{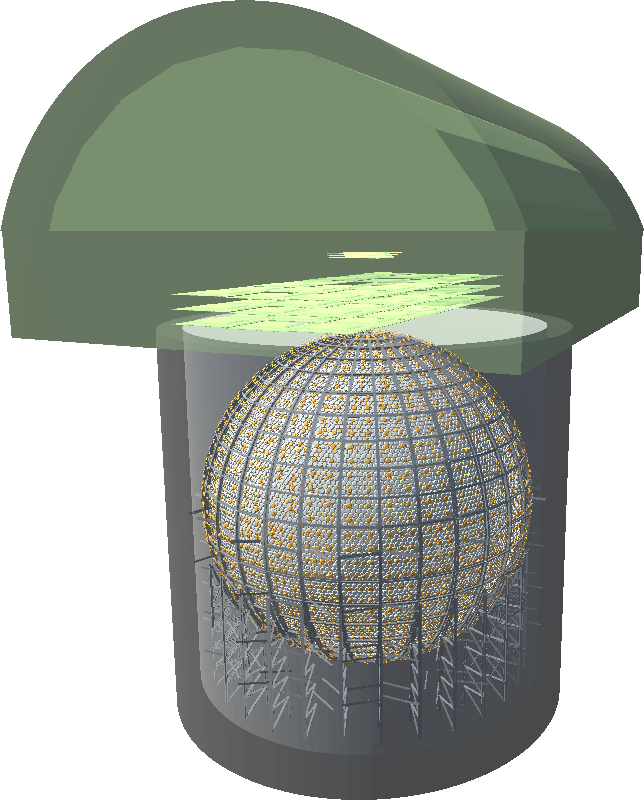}
  }
  \hfill
  \subfloat[EicC detector]{
        \label{fig:EicC}
        \includegraphics[width=0.5\textwidth]{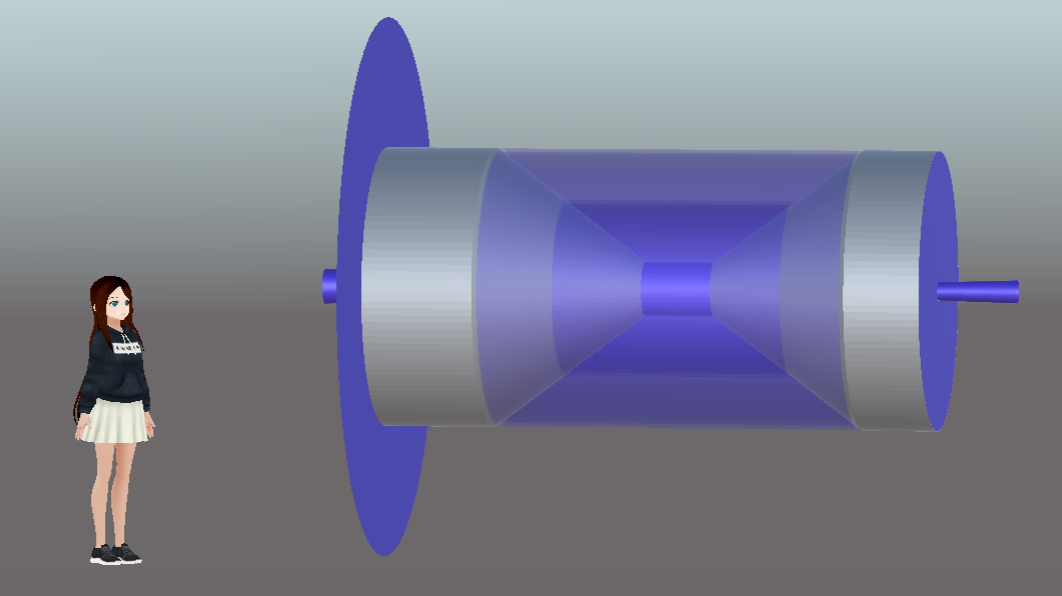}
  }
  \caption{JUNO and EicC detector display in Unity. A life-sized human model in (b) serves as a scale reference for EicC detector.}
  \label{fig:all}
\end{figure}

The Jiangmen Underground Neutrino Observatory~(JUNO) is a neutrino experiment~\cite{JUNO SW}.
Its detector system includes a Central Detector~(CD) containing liquid scintillator, a water Cherenkov detector, and a Top Tracker~(TT). 
The CD is housed within a water pool fitted with photomultiplier tubes~(PMTs) to capture Cherenkov light from cosmic muons, while the TT is composed of plastic scintillator arrays positioned above the pool. 
After converting the JUNO detector description from Geant4 to FBX, we imported it into Unity for visualization, as illustrated in Figure~\ref{fig:JUNO}.

\subsection{From ROOT to Unity with EicC detector}
\label{subsec:EicC}
The Electron-Ion Collider in China (EicC) is a proposed HEP facility currently in the research and development phase~\cite{EicC}, with its detector description originally constructed using ROOT. This description was first translated into GDML format and then processed through our interface to produce an FBX file, which is presented in Figure~\ref{fig:EicC}.


\subsection{From GDML to Unity with BESIII detector}
\label{subsec:BESIII}
\begin{figure}[ht]
  \centering
  \subfloat[BESIII detector]{
        \label{fig:BESIII}
        \includegraphics[width=0.44\textwidth]{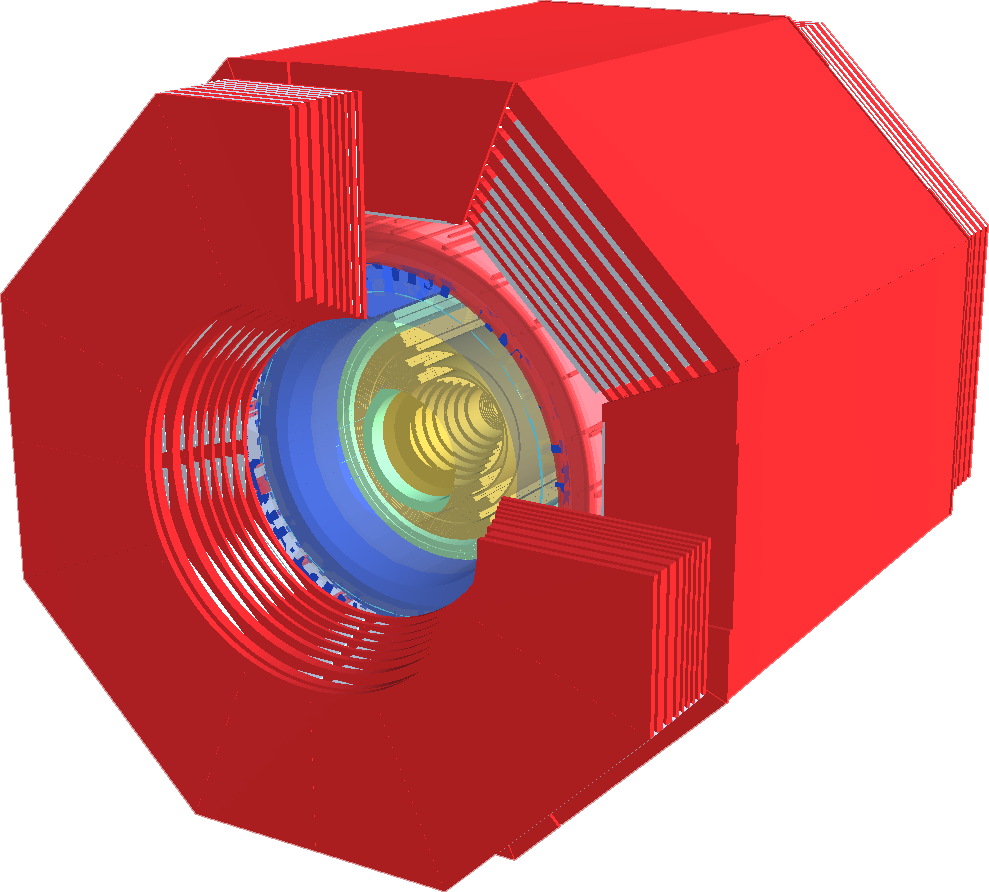}
  }
  \hfill
  \subfloat[CEPC detector]{
        \label{fig:CEPC}
        \includegraphics[width=0.44\textwidth]{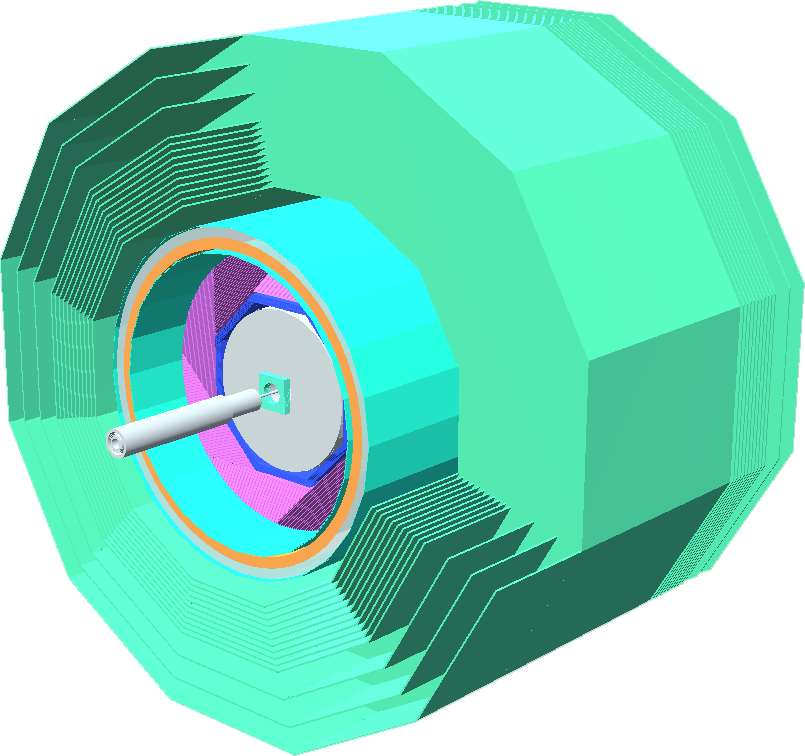} 
  }  
  \caption{BESIII and CEPC detector display in Unity.}
  \label{fig:BESIII&CEPC}
\end{figure}

The Beijing Spectrometer Experiment (BESIII) at BEPCII is designed for precise studies in $\tau$-charm physics~\cite{BESIII description,besiiidata}. 
Its detector features a multilayer drift chamber (MDC), time-of-flight system (TOF), electromagnetic calorimeter (EMC), superconducting solenoidal magnet, and muon counter (MUC). 
Using our interface, the GDML-based description of BESIII was transformed into FBX format and rendered in Unity, as shown in Figure~\ref{fig:BESIII}. 

\subsection{From DDD4hep to Unity with CEPC detector}
\label{subsec:CEPC}
The Circular Electron-Positron Collider (CEPC) is an electron-positron collider under development. 
Its detector comprises a vertex detector (VXD), forward and endcap tracking detectors (FTD and ETD), time projection chamber (TPC), electromagnetic calorimeter (ECAL), hadronic calorimeter (HCAL), a 3 Tesla solenoid, and a muon detection system (MUD) integrated into the return yoke. 
The detector description~\cite{CEPC}, provided in DD4hep, was converted to FBX and visualized in Unity, as depicted in Figure~\ref{fig:CEPC}.

\section{Summary}
\label{sec:summary}
In HEP experiments, detector descriptions are typically developed using specialized formats such as Geant4, ROOT, GDML, and DD4hep, 
while these formats lack compatibility with widely adopted 3D modeling standard format like FBX. 
This disparity obstructs the direct import of detector models into industrial software, limiting their application in advanced HEP workflows. 
To overcome this challenge, we propose an interface that facilitates the conversion of all four common HEP detector description formats into FBX files. 
This innovation supports the development of sophisticated HEP applications using industrial tools like Unity, enabling enhancements in areas such as detector design, event visualization, data monitoring, and related domains.

\section*{Acknowledgements}
We thank the helpful discussions in the offline software group of BESIII and JUNO collaboration, and thank the colleagues from IHEP, CAS working on CEPC, and colleagues from IMP, CAS working on EicC, for providing the detector description data files. 
This work is supported by the National Natural Science Foundation of China (Grant Nos. 12175321, W2443004, 11975021, 11675275, and U1932101), National Key Research and Development Program of China (Nos. 2023YFA1606000 and 2020YFA0406400), Strategic Priority Research Program of the Chinese Academy of Sciences (No. XDA10010900), National College Students Science and Technology Innovation Project, and Undergraduate Base Scientific Research Project of Sun Yat-sen University.

\end{document}